\documentclass[letter]{aa}  
\usepackage{graphicx}
\usepackage{txfonts}
\usepackage{hyperref}
\usepackage{soul}

\usepackage{xcolor}
\usepackage[normalem]{ulem}
\usepackage{orcidlink}
\usepackage{threeparttable} 

\usepackage{subfigure}

\begin{document} 

\title{The convective-transition gap in 47\,Tucanae with JWST}

\author{M.\,Scalco\inst{1}\orcidlink{0000-0001-8834-3734}\fnmsep\thanks{\email{mscalco@iu.edu}}
\and 
L.\,R.\,Bedin\inst{2}\orcidlink{0000-0003-4080-6466}
\and
M.\,Libralato\inst{2}\orcidlink{0000-0001-9673-7397}
\and
M.\,Griggio\inst{3}\orcidlink{0000-0002-5060-1379}
\and 
E.\,Vesperini\inst{1}\orcidlink{0000-0003-2742-6872}
\and 
R.\,Gerasimov\inst{4}\orcidlink{0000-0003-0398-639X}
}

\institute{
Department of Astronomy, Indiana University, Swain West, 727 E. 3rd Street, Bloomington, IN 47405, USA
\and
Istituto Nazionale di Astrofisica, Osservatorio Astronomico di Padova, Vicolo dell'Osservatorio 5, Padova I-35122, Italy
\and 
Space Telescope Science Institute, 3700 San Martin Dr, Baltimore, MD 21218, USA
\and
Department of Physics and Astronomy, University of Notre Dame, Nieuwland Science Hall, Notre Dame, IN 46556, USA
}

\date{XXX,YYY,ZZZ}
 
\abstract
{We report the detection of the convective-transition gap along the lower main sequence (MS) of the Galactic globular cluster (GC) 47\,Tucanae using deep \textit{James Webb} Space Telescope (JWST) observations. We analyse the stellar luminosity function (LF) in the vicinity of the gap location using two independent colour--magnitude diagrams (CMDs). In both cases, the LF is characterised by a narrow gap followed by a discontinuous increase in the stellar number counts, and presents an overall morphology qualitatively consistent with that reported in previous studies. By modelling the observed LFs with an empirical likelihood model, we estimate the statistical significance of the observed feature. Additional JWST passbands exhibit similar signatures, providing further evidence for the presence of a discontinuity.} 


\titlerunning{The convective-transition gap in 47\,Tuc}
\authorrunning{Scalco et al.}
\maketitle

\section{Introduction}\label{Section1}

A narrow underdensity of stars along the lower main sequence (MS), commonly referred to as the \textit{convective-transition gap}, \textit{M-dwarf gap}, or \textit{Gaia gap}, was first identified in nearby field stars using \textit{Gaia} photometry by \citet{2018ApJ...861L..11J} and later investigated using both observational and theoretical studies \citep[e.g.][]{2018MNRAS.480.1711M,2020AJ....160..102J,2021ApJ...907...53F,2021A&A...650A.184M,2023ApJ...944..129B}. The feature is thought to be associated with the transition between partially and fully convective stellar interiors and the complex interplay between convection and $^3$He production and mixing in low-mass stars \citep{convective_kissing,2018A&A...619A.177B}. Evidence for the same phenomenon was also identified in 2MASS photometry, although at lower statistical significance \citep{2018ApJ...861L..11J,2020AJ....160..102J}.\looseness=-4

The detection of the convective-transition gap in Galactic globular clusters (GCs) has proven considerably more challenging, not only because of crowding, but also because --- even in the nearest GCs --- the intrinsic faintness of the stars involved makes it difficult to achieve the photometric precision required to resolve such a narrow feature. Furthermore, even with infinite photometric precision and arbitrarily deep observations, significant sampling noise remains an important limitation, requiring large stellar samples to detect the feature with high statistical confidence. As a consequence, populous GCs observed over wide fields are required to investigate this feature.\looseness=-4

Recently, \citet{2026A&A...711A..88G,2026arXiv260628504L} reported the first statistically significant detection of the feature in the GCs NGC\,6397 and NGC\,104 using \textit{Euclid} observations. The identification of the gap in GCs is particularly important because these systems provide cleaner stellar populations than the heterogeneous field-star samples available in the solar neighbourhood and therefore offer a more direct comparison with stellar evolution models.\looseness=-4

Independent confirmations using different instruments and data-reduction strategies are essential to establish the robustness of this result and to assess the ubiquity of the convective-transition gap in old stellar populations. In this Letter we investigate the lower MS of 47\,Tucanae (or NGC\,104, hereafter 47\,Tuc) using deep \textit{James Webb} Space Telescope (JWST) observations.\footnote{At the time of submission, we became aware of a preprint by Milone et al. (2026, arXiv:2606.20979), containing a similar investigation. Our work was conducted independently and offers a specifically targeted, in-depth analysis of this phenomenon employing multiple datasets.}\looseness=-4

\section{Data reduction and analysis}\label{Section2}
The data analysed in this work are based on a combination of JWST Near Infrared Camera \citep[NIRCam;][]{2023PASP..135b8001R} observations obtained as part of several programmes targeting 47\,Tuc. The dataset comprises imaging acquired through multiple wide- and ultra-wide-band filters. A summary of the observations used in this work is provided in Table\,\ref{table:dataset}, while their field of view (FOV) is shown in Fig.\,\ref{FOV}. The observations and data reduction procedures adopted for part of the dataset analysed here were presented by \citet{2025A&A...694A..68S}. The remaining observations were reduced following the same methodology. Briefly, the \texttt{\_uncal} exposures were processed using a development version of the JWST pipeline to produce \texttt{\_cal} images. Stellar photometry and astrometry were obtained from the \texttt{\_cal} images through a two-step procedure consisting of a first and a second stage. Photometry was then calibrated onto the VEGA-mag system. We refer to \citet{2025A&A...694A..68S} for a detailed description of the reduction procedures. We note, however, that unlike in \citet{2025A&A...694A..68S}, no differential reddening or zero-point variation corrections were applied in the present work, as such corrections could potentially smooth out the subtle feature investigated here. We make the photometric catalogue publicly available as supplementary online material to this paper and through our website\footnote{\url{https://web.oapd.inaf.it/bedin/files/PAPERs_eMATERIALs/JWST/Paper_47Tuc_MdwarfGap/}}.\looseness=-4

\begin{figure}
\includegraphics[width=\columnwidth]{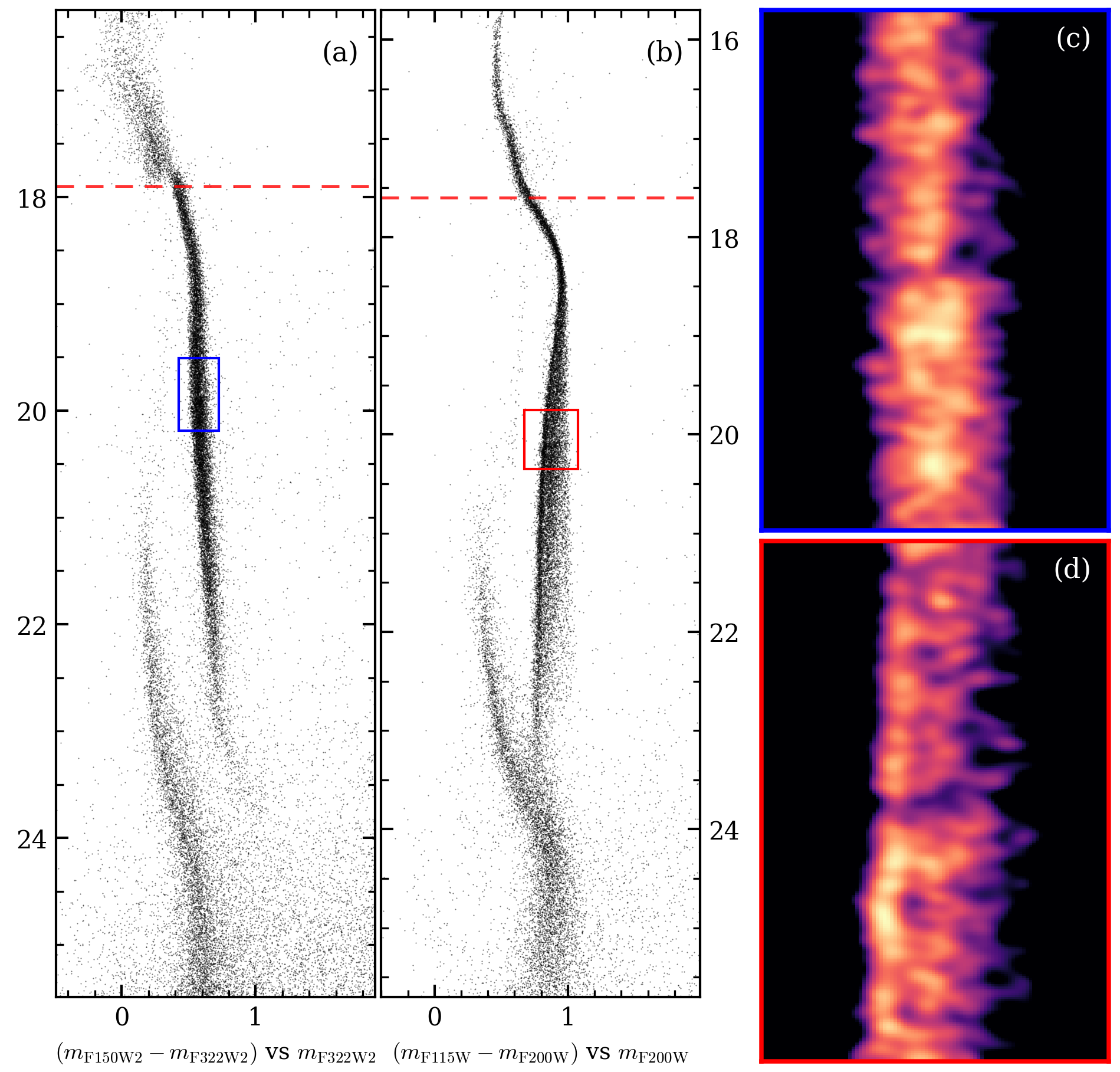}
 \caption{The relevant CMDs where the convective gap is clearly detected. (a) $m_{\rm F322W2}$ versus $m_{\rm F150W2}-m_{\rm F322W2}$ CMD as presented in \citet{2025A&A...694A..68S}. (b) $m_{\rm F200W}$ versus $m_{\rm F115W}-m_{\rm F200W}$ CMD. The dashed red lines mark the approximate saturation limit. (c)-(d) Hess diagrams of the regions enclosed by the blue and red boxes in (a) and (b), respectively.}
 \label{fig1}
\end{figure}

Figure\,\ref{fig1} presents two colour-magnitude diagrams (CMDs): $m_{\rm F322W2}$ versus $m_{\rm F150W2}-m_{\rm F322W2}$ in panel\,(a), and $m_{\rm F200W}$ versus $m_{\rm F115W}-m_{\rm F200W}$ in panel\,(b). As shown in \citet{2026A&A...711A..88G}, the convective-transition gap is expected to appear as a narrow local underdensity along the MS, followed by a discontinuous change (or step) in the stellar number counts. Scalco et al.\,(2026, accepted) estimated that, in 47\,Tuc and in the F322W2 filter, the instability region associated with the transition between convective stellar interiors and radiative layers should span a region of $\sim$0.02--0.03 mag and be located around $19.79 \lesssim m_{\rm F322W2} \lesssim 19.94$ (corresponding to a mass range of $0.331 \lesssim M/M_{\odot} \lesssim 0.337$), depending on the adopted model. Panel\,(c) of Fig.\,\ref{fig1} shows a zoomed-in Hess diagram of the CMD in panel\,(a) around this magnitude range. The diagram reveals a subtle but discernible underdensity of stars. A similar feature is also visible in the CMD of panel\,(b), as shown in the corresponding zoomed-in Hess diagram (panel\,d).\looseness=-4

\begin{figure}
 \includegraphics[width=\columnwidth]{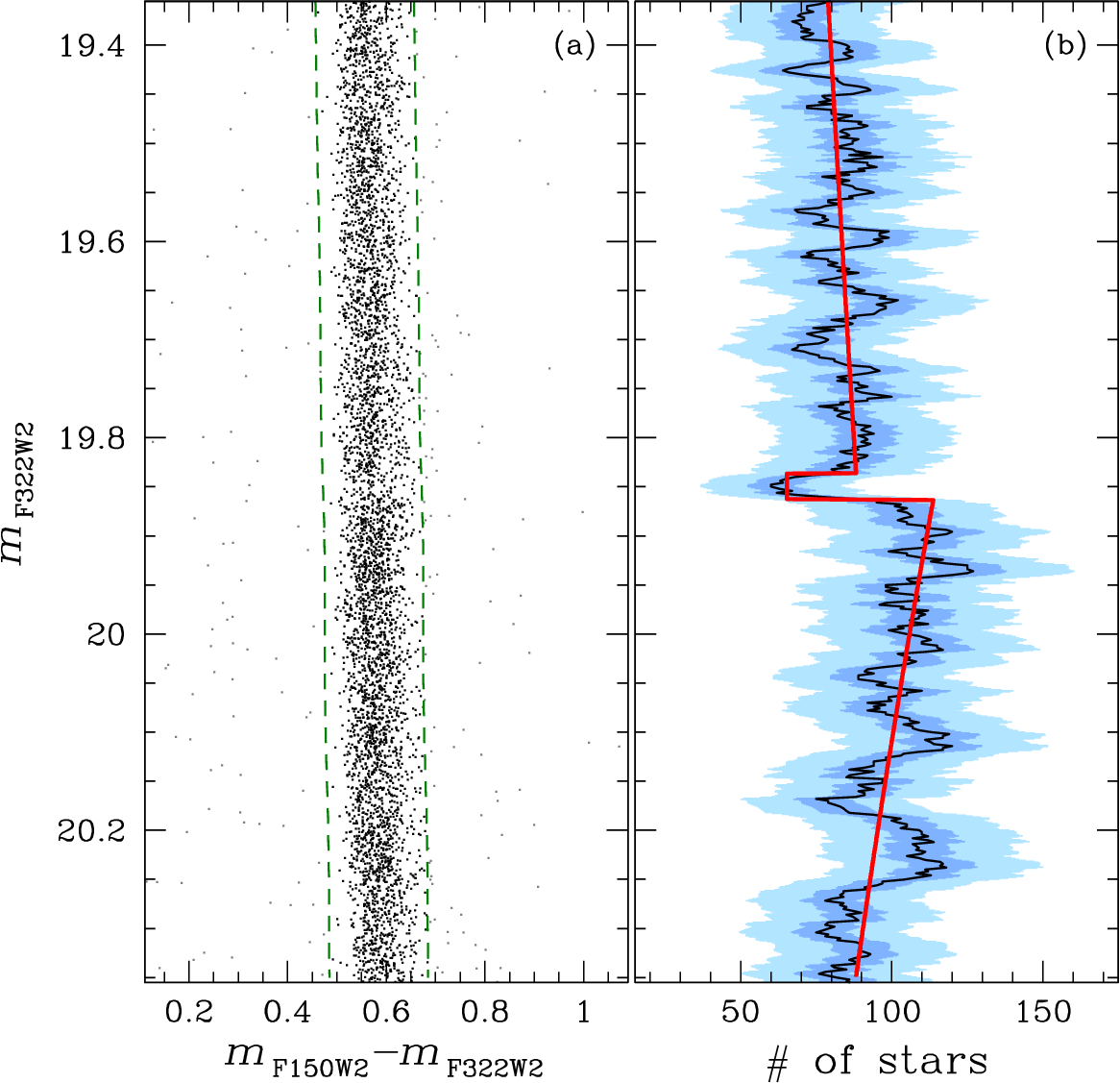}
 \caption{(a) $m_{\rm F322W2}$ versus $m_{\rm F150W2}-m_{\rm F322W2}$ CMD of 47\,Tuc around the expected location of the convective gap. The green dashed lines mark the colour selection adopted to construct the LF. (b) Luminosity function of the selected stars as a function of $m_{\rm F322W2}$ (black line). The light- and dark-blue shaded regions indicate the corresponding 1$\sigma$ and 3$\sigma$ confidence intervals, respectively. The best-fitting empirical likelihood model is shown in red.}
 \label{fig:lf1}
\end{figure}

\begin{figure}
 \includegraphics[width=\columnwidth]{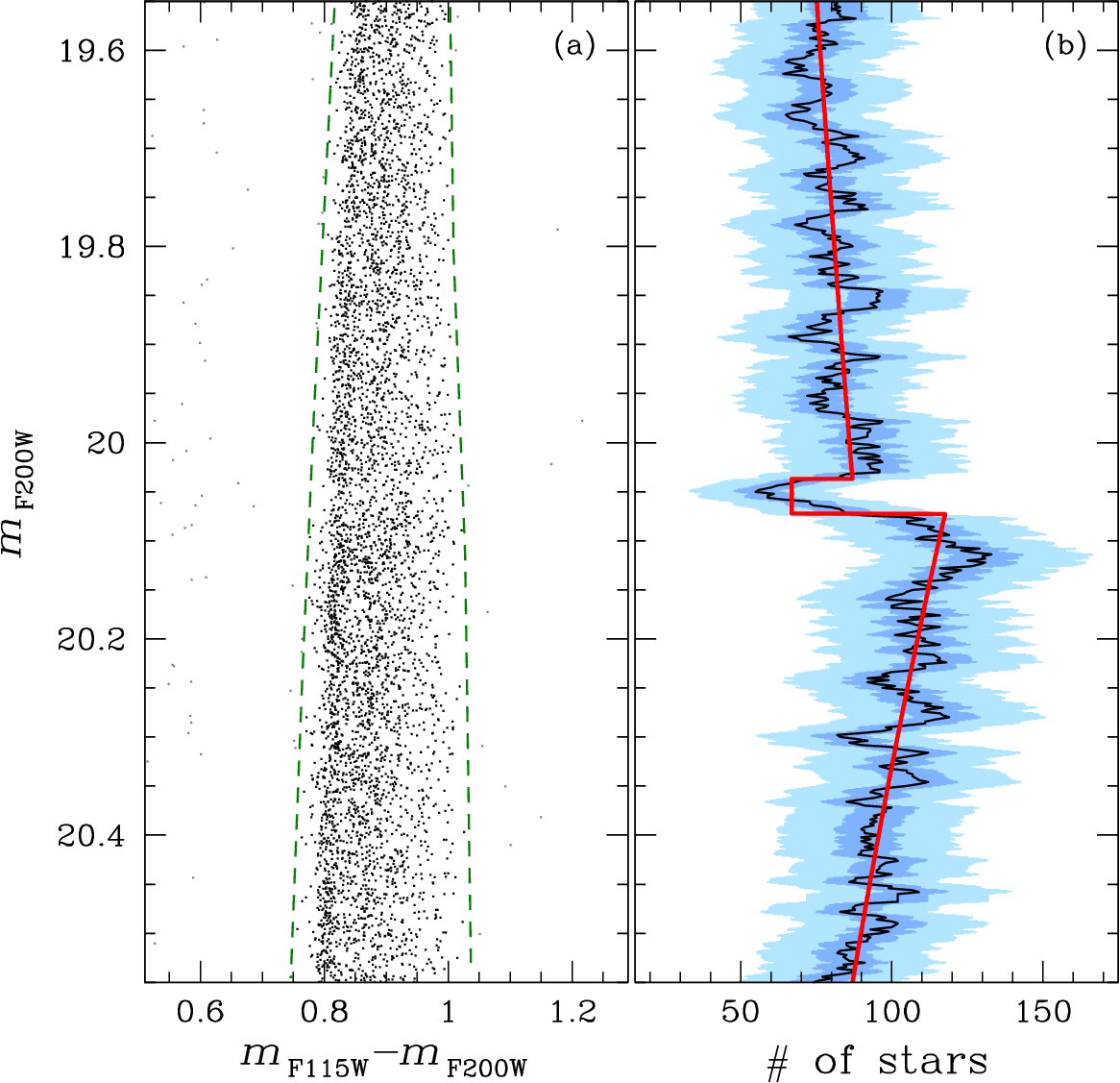}
 \caption{Same as Fig.\,\ref{fig:lf1} but for the $m_{\rm F200W}$ versus $m_{\rm F115W}-m_{\rm F200W}$ CMD.}
 \label{fig:lf2}
\end{figure}

To investigate the morphology of this region of the MS, we constructed the luminosity function (LF) of the stars around this feature. We first selected a sample of likely cluster members by defining by hand two fiducial lines enclosing the MS in the region of interest (dashed green lines in panels\,(a) of Figs.\,\ref{fig:lf1} and \ref{fig:lf2}). The adopted fiducials were chosen as a compromise between maximizing the number of MS and minimizing contamination from binaries and field sources.\looseness=-4

We note that no proper-motion (PM) selection was applied in this analysis, as PM measurements are available only for a fraction of the observed field. Restricting the sample to stars with PM information would therefore significantly reduce the number of stars and potentially diminish the visibility of the expected gap. Contamination from the Small Magellanic Cloud (SMC) is nevertheless expected to be minimal within the adopted selection region, since the SMC MS lies systematically bluer than the 47\,Tuc MS in this magnitude range (see Fig.\,\ref{fig1}) and therefore falls largely outside our selection boundaries. In addition, Galactic field stars are expected to be relatively sparsely and smoothly distributed in the CMD.\looseness=-4

To minimise the dependence on the arbitrary placement of the bin boundaries, we computed a set of ten LFs using a bin width of 0.02\,mag and successive shifts of 0.002\,mag (i.e., one tenth of the bin width). The resulting measurements were combined to derive a continuous estimate of the LF. The derived LFs are shown in panels\,(b) of Figs.\,\ref{fig:lf1} and \ref{fig:lf2} as a black curve. The light- and dark-blue shaded regions indicate the corresponding 1$\sigma$ and 3$\sigma$ confidence intervals, respectively. The smoothed LFs show a local minimum visible at $m_{\rm F322W2}\simeq19.85$ and $m_{\rm F200W}\simeq20.05$. In both LFs, the deficit of stars at the gap location is unlikely to arise from a statistical fluctuation, as the local minimum remains visible even within a 3$\sigma$ confidence interval. The morphology of the two LFs is similar and qualitatively consistent with that reported by \citet{2026A&A...711A..88G}: starting from the bright end of the LF, the stellar counts increase towards fainter magnitudes, reach a narrow minimum, followed by a step-like increase to higher values. Beyond this discontinuity, the stellar counts decrease towards fainter magnitudes.\looseness=-4

To infer the properties of the observed gaps and estimate their statistical significance, we modelled the LFs in the vicinity of the gaps using an empirical likelihood model, following an approach similar to that adopted by \citet{2026A&A...711A..88G}. The analysis was performed independently for the stars selected in the two CMDs shown in Figs.\,\ref{fig:lf1} and \ref{fig:lf2}. The fit was performed directly on the individual stellar magnitudes, avoiding any dependence on the adopted binning scheme. The stellar number counts were modelled as a function of the magnitude $m$, as
\begin{equation}
\phi(m) = e^{a(m-m_{\rm ref}) + b},
\end{equation}
where $a$ and $b$ represent the slope and intercept of the LF, respectively, and $m_{\rm ref}$ is a fixed reference magnitude. The gap was modelled as a flat depression centred at magnitude $m_0$ and spanning a width $w$. The gap boundaries are defined as
\begin{equation}
m_{\rm min}=m_0-\frac{w}{2},
\qquad
m_{\rm max}=m_0+\frac{w}{2}.
\end{equation}
The depth of the gap was parametrised by a multiplicative factor $D$, such that $D=1$ corresponds to a gap-free model and $D<1$ indicates a deficit of stars within the gap.\looseness=-4

To account for the discontinuity in the LF at the faint edge of the gap, we introduced a step parameter $R$, defined as the ratio between the LF immediately above and below the gap. We allowed the LF slope to change beyond the gap through a parameter $\Delta a$, such that the faint-side LF follows a slope $a+\Delta a$. The complete model can therefore be written as
\begin{equation}
\phi(m)=
\begin{cases}
e^{\,a(m-m_{\rm ref}) + b},
&
m < m_{\rm min}, 
\\
D\phi_{\rm gap},
&
m_{\rm min}\le m \le m_{\rm max},
\\
R e^{(a+\Delta a)(m-m_{\rm max})}
e^{a(m_{\rm max}-m_{\rm ref})+b},
&
m > m_{\rm max},
\end{cases}
\end{equation}
where $\phi_{\rm gap}$ is a constant level within the gap, chosen to be equal to the average of the LF values at the two gap boundaries.\looseness=-4

The model contains seven free parameters: the LF slope and normalisation ($a,b$), the gap position and width ($m_0,w$), the gap depth ($D$), the amplitude of the step ($R$), and the change in LF slope beyond the gap ($\Delta a$). We estimated their best-fit values and uncertainties by sampling the parameter space with a Markov chain Monte Carlo (MCMC) approach, using the Goodman–Weare algorithm \citep{MCMC} implemented in the \texttt{emcee} Python package \citep{emcee_2013}. We employed 32 MCMC walkers and 5\,000 steps per walker. Parameter estimates were taken as the median of the posterior distributions, while the uncertainties correspond to the 16th and 84th percentiles of the posterior samples. The best-fitting models for the two LFs are shown in panel\,(b) of Figs.\,\ref{fig:lf1} and \ref{fig:lf2}.\looseness=-4

To assess the statistical significance of the gaps, we compared the logarithmic likelihood of the best-fit model with that of an otherwise identical gap-free model (D=1). The observed difference in likelihood for the $m_{\rm F322W2}$ ($m_{\rm F200W}$) magnitude $\Delta \ln \mathcal{L} = \ln \mathcal{L}_{\rm gap} - \ln \mathcal{L}_{\rm no\,gap} = 9.92$ ($9.77$), was compared to its expected distribution, estimated from 1000 synthetic stellar samples with the same size as the observed dataset. To establish a baseline, we also generated 1000 synthetic datasets from the gap-free likelihood. We inferred expected distributions of $\Delta \ln \mathcal{L} = 9.37 \pm 4.15$ ($12.78 \pm 4.68$) and $\Delta \ln \mathcal{L} = -11.43 \pm 5.00$ ($-11.44 \pm 5.27$) for the best-fit and gap-free models, respectively. The observed value of $\Delta \ln \mathcal{L}$ is fully consistent with the distribution predicted by the best-fit model. In contrast, the observed LF lies approximately $4.4\sigma$ ($4.0\sigma$) away from the $\Delta \ln \mathcal{L}$ distribution predicted by the gap-free model. Dedicated artificial-star (AS) tests are presented in Appendix\,\ref{Appendix:E}. \looseness=-4

Figures\,\ref{posterior322} and \ref{posterior200} show the posterior distributions of the four parameters describing the gap for the $m_{\rm F322W2}$ and $m_{\rm F200W}$, respectively. The posterior distributions of the gap depth ($D$) and step amplitude ($R$) are approximately Gaussian in both cases. The gap centre ($m_0$) is also well constrained, although the $m_{\rm F322W2}$ analysis shows a mild asymmetry towards brighter magnitudes. The width parameter ($w$) for the $m_{\rm F322W2}$ displays an asymmetric distribution characterized by a sharp maximum at $w\simeq0.015$\,mag. The best-fit parameters obtained from the $m_{\it F322W2}$ and $m_{\it F200W}$ magnitudes are presented in Tables\,\ref{tab:gap_fit}. The inferred gap properties are remarkably consistent between the two analyses. In both cases, the parameters $D$ and $R$ differ from unity by between 3-4 standard deviations, suggesting that both the gap and the adjacent step in star counts are detected at high statistical confidence.\looseness=-4

\begin{table}
\centering
\caption{Best-fit parameters of the empirical LF model analyses.}
\renewcommand{\arraystretch}{1.4}
\begin{tabular*}{\columnwidth}{@{\extracolsep{\fill}}lccc}
\hline
\hline
Parameter & $m_{\rm F322W2}$ & $m_{\rm F200W}$ & Units \\
\hline
$a$ &
$0.2297_{-0.1696}^{+0.1674}$ &
$0.2986_{-0.1592}^{+0.1710}$ &
mag$^{-1}$ \\

$b$ &
$8.4055_{-0.0551}^{+0.0546}$ &
$8.3971_{-0.0543}^{+0.0574}$ &
-- \\

$m_0$ &
$19.8499_{-0.0056}^{+0.0042}$ &
$20.0548_{-0.0041}^{+0.0043}$ &
mag \\

$w$ &
$0.0268_{-0.0115}^{+0.0134}$ &
$0.0353_{-0.0087}^{+0.0118}$ &
mag \\

$D$ &
$0.6462_{-0.1161}^{+0.1160}$ &
$0.6540_{-0.0865}^{+0.0922}$ &
-- \\

$R$ &
$1.2819_{-0.0801}^{+0.0848}$ &
$1.3392_{-0.0815}^{+0.0944}$ &
-- \\

$\Delta a$ &
$-0.7537_{-0.2295}^{+0.2166}$ &
$-0.9302_{-0.2234}^{+0.2220}$ &
mag$^{-1}$ \\
\hline
\end{tabular*}
\label{tab:gap_fit}
\end{table}
For completeness, Fig.\,\ref{ist_tot} shows the LFs derived from the remaining JWST/NIRCam filters. Although less statistically significant, several passbands exhibit features qualitatively consistent with those identified in the F322W2 and F200W analyses, providing additional support for the presence of a discontinuity.\looseness=-4

Finally, we compare our results with the recent findings by \citet{2026arXiv260628504L}. To enable a direct comparison, we transformed the magnitudes at which the convective-transition gap is located in the JWST filters, as measured in this work, into the \textit{Euclid} $I_{\rm E}$ band. We adopted the same BaSTI-IAC isochrones \citep{2021ApJ...908..102P}, distance modulus ($m-M=13.24$), reddening ($E(B-V)=0.04$), and extinction coefficients as \citet{2026arXiv260628504L}. We also applied an empirical offset of 0.183\,mag in the $I_{\rm E}$ band as reported by \citet{2026arXiv260628504L}. The gap magnitudes ($m_0$) reported in Table\,\ref{tab:gap_fit} correspond to $I_{\rm E}=22.87$ and $I_{\rm E}=22.92$ for the $m_{\rm F322W2}$ and $m_{\rm F200W}$ filters, respectively. These values are in good agreement with the \textit{Euclid} measurement of $I_{\rm E}=22.921^{+0.067}_{-0.014}$ reported by \citet{2026arXiv260628504L}.\looseness=-4

\section{Multiple stellar populations}
\label{sec:mpops}

It is well established that 47\,Tuc hosts multiple stellar populations characterized by variations in their light-element abundances \citep[e.g.,][]{2009ApJ...697L..58A,2013ApJ...771L..15R}. These stars can be broadly divided into two main components, commonly referred to as first- and second-population (1P and 2P) stars. To investigate the sensitivity of the convective-transition gap to these chemical variations, we separated the two populations using a two-pseudo-colour diagram (TpCD) constructed from the JWST colours $m_{\rm F115W}-m_{\rm F200W}$ and $m_{\rm F150W2}-m_{\rm F322W2}$. The TpCD was built following the procedure adopted in similar studies \citep[see][and references therein]{2024A&A...688A.180S,2024A&A...689A..59S}. Briefly, the two colours were verticalised with respect to fiducial lines tracing the MS and then combined into a single diagram. We defined by hand empirical boundaries enclosing the two main overdensities visible in the TpCD and used them to isolate the 1P and 2P samples (see Fig.\,\ref{tpcd} for details). To obtain clean samples of the two populations, we restricted the analysis to stars with high-quality photometry and cluster membership established through PM selection. Together with the requirement that stars be present in overlapping datasets, this significantly reduces the number of stars available for this analysis compared to the gap search described in the previous sections.\looseness=-4

\begin{figure}
    \centering
    \includegraphics[width=\columnwidth]{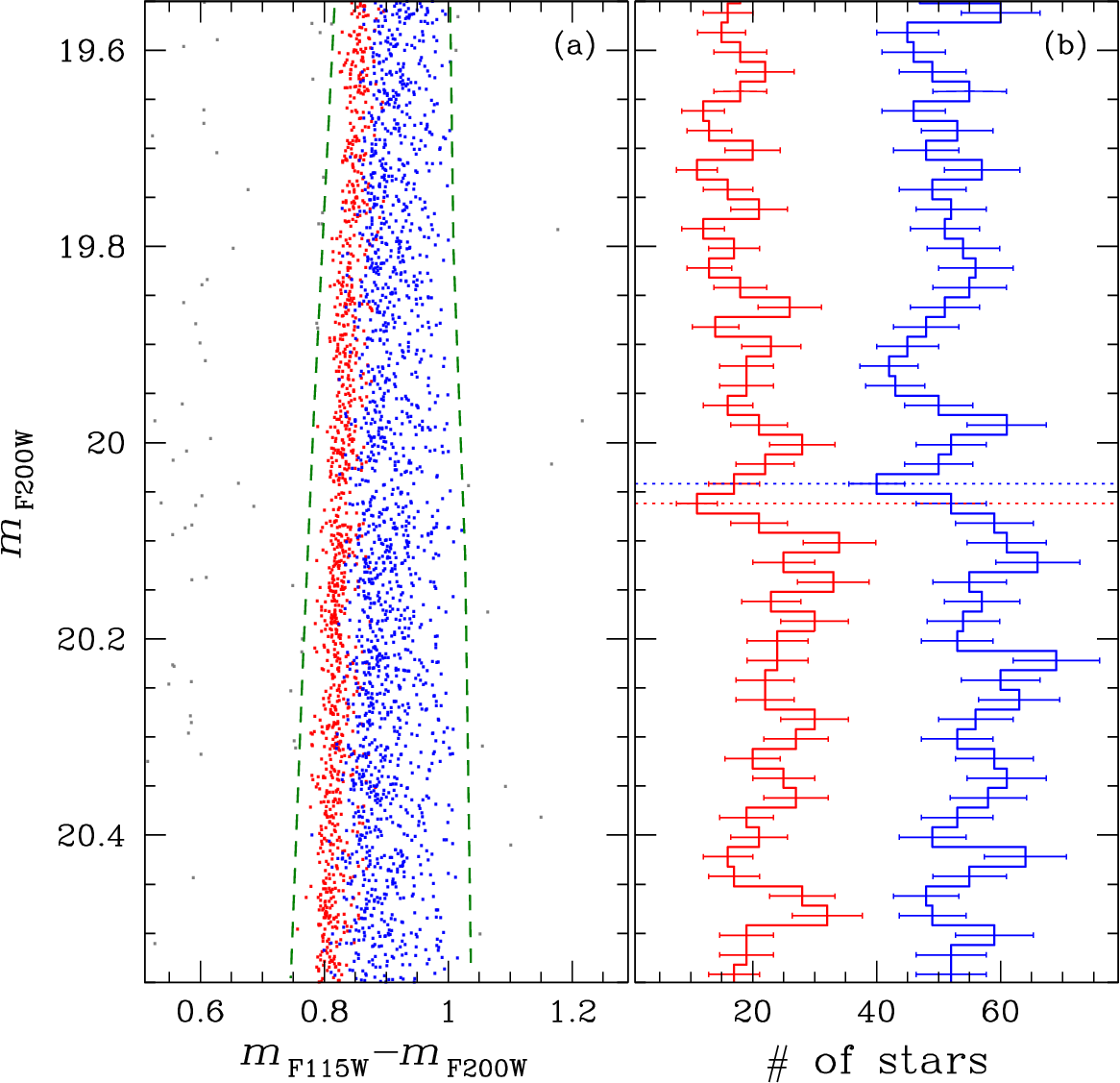}    
    \caption{The convective-transition gap for the 1P and 2P stars. (a) $m_{\rm F200W}$ versus $m_{\rm F115W}-m_{\rm F200W}$ CMD with stars belonging to the 1P and 2P shown in red and blue, respectively. (b) Corresponding LFs of the two populations; the LF of the second population (blue) has been shifted horizontally for clarity. The dotted horizontal lines mark the tentative locations of the convective-transition gap for both sub-samples, illustrating the hint of a shift towards brighter magnitudes for the 2P.}
    \label{fig:mpop_gap}
\end{figure}

Despite the substantial reduction in sample size resulting from both the population separation and the selection criteria, we investigated whether the convective-transition gap could be identified independently within each population. As shown in Fig.\,\ref{fig:mpop_gap} the gap signature appears to be present in both populations. The comparison between the two LFs provides a tentative indication that the gap may occur at slightly brighter magnitudes in the 2P than in the 1P.\looseness=-4

The presence of a shift between the two populations would be qualitatively consistent with theoretical expectations, since the abundance variations characterising the 2P (e.g. oxygen depletion and modest helium enhancement) are expected to affect atmospheric opacities and stellar structure, thereby modifying the luminosity at which the convective transition occurs. This may also contribute to the different visibility of the gap among the various JWST filter combinations (see Fig.\,\ref{ist_tot}). If the gap is inclined with respect to the magnitude axis in a given CMD, its signature is spread over a broader magnitude interval, reducing the contrast of the corresponding LF and making the feature more difficult to detect. However, given the limited number of stars available in each sub-sample, the statistical significance of this possible offset remains low. A robust determination of any difference in the location of the convective-transition gap between the two populations will require a substantially larger stellar sample.\looseness=-4

\section{Summary and Conclusions}\label{S&C}

We identified an under-density of stars within 47\,Tuc at magnitudes consistent with the expected location of the convective-transition gap and quantified its statistical significance through an analysis of the stellar LF. Our results provide an independent assessment of the presence of the convective-transition gap in one of the best-studied GCs. To date, the convective-transition gap has been identified in only two GCs, both using \textit{Euclid} observations. The present JWST detection therefore provides an independent confirmation of this feature using a different telescope, photometric system, and data-reduction strategy.\looseness=-4

Going beyond this initial detection, accurate measurements of the magnitudes and amplitudes of the gap observed across multiple JWST passbands will provide crucial empirical constraints. These multi-band observations will help theorists develop detailed physical models of this feature. Ultimately, the convective-transition gap will serve as a powerful new diagnostic tool to calibrate low-mass stellar interiors, as well as to probe the chemical compositions and distributions of multiple stellar populations across GCs with different heavy-element abundances.\looseness=-4

\begin{acknowledgements}
MS and EV acknowledge support from STScI funding associated with JWST program GO-5896. L.R. Bedin, M. Libralato acknowledge support by INAF under the WFAP project, f.o.:1.05.23.05.05.
\end{acknowledgements}

\bibliographystyle{aa}
\bibliography{main.bib}

@ARTICLE{2010arXiv1012.3224H,
       author = {{Harris}, William E.},
        title = "{A New Catalog of Globular Clusters in the Milky Way}",
      journal = {arXiv e-prints},
         year = 2010,
        month = dec,
          eid = {arXiv:1012.3224},
        pages = {arXiv:1012.3224},
          doi = {10.48550/arXiv.1012.3224},
archivePrefix = {arXiv},
       eprint = {1012.3224},
 primaryClass = {astro-ph.GA},
       adsurl = {https://ui.adsabs.harvard.edu/abs/2010arXiv1012.3224H}
}

@ARTICLE{2024A&A...689A..59S,
       author = {{Scalco}, M. and {Libralato}, M. and {Gerasimov}, R. and {Bedin}, L.~R. and {Vesperini}, E. and {Nardiello}, D. and {Bellini}, A. and {Griggio}, M. and {Apai}, D. and {Salaris}, M. and {Burgasser}, A. and {Anderson}, J.},
        title = "{JWST imaging of the closest globular clusters: III. Multiple populations along the low-mass main-sequence stars of NGC 6397}",
      journal = {\aap},
         year = 2024,
        month = sep,
       volume = {689},
          eid = {A59},
        pages = {A59},
          doi = {10.1051/0004-6361/202450589},
archivePrefix = {arXiv},
       eprint = {2405.01635},
 primaryClass = {astro-ph.SR},
       adsurl = {https://ui.adsabs.harvard.edu/abs/2024A&A...689A..59S}
}

@ARTICLE{2023PASP..135b8001R,
       author = {{Rieke}, Marcia J. and {Kelly}, Douglas M. and {Misselt}, Karl and {Stansberry}, John and {Boyer}, Martha and {Beatty}, Thomas and {Egami}, Eiichi and {Florian}, Michael and {Greene}, Thomas P. and {Hainline}, Kevin and {Leisenring}, Jarron and {Roellig}, Thomas and {Schlawin}, Everett and {Sun}, Fengwu and {Tinnin}, Lee and {Williams}, Christina C. and {Willmer}, Christopher N.~A. and {Wilson}, Debra and {Clark}, Charles R. and {Rohrbach}, Scott and {Brooks}, Brian and {Canipe}, Alicia and {Correnti}, Matteo and {DiFelice}, Audrey and {Gennaro}, Mario and {Girard}, Julien H. and {Hartig}, George and {Hilbert}, Bryan and {Koekemoer}, Anton M. and {Nikolov}, Nikolay K. and {Pirzkal}, Norbert and {Rest}, Armin and {Robberto}, Massimo and {Sunnquist}, Ben and {Telfer}, Randal and {Wu}, Chi Rai and {Ferry}, Malcolm and {Lewis}, Dan and {Baum}, Stefi and {Beichman}, Charles and {Doyon}, Ren{\'e} and {Dressler}, Alan and {Eisenstein}, Daniel J. and {Ferrarese}, Laura and {Hodapp}, Klaus and {Horner}, Scott and {Jaffe}, Daniel T. and {Johnstone}, Doug and {Krist}, John and {Martin}, Peter and {McCarthy}, Donald W. and {Meyer}, Michael and {Rieke}, George H. and {Trauger}, John and {Young}, Erick T.},
        title = "{Performance of NIRCam on JWST in Flight}",
      journal = {\pasp},
         year = 2023,
        month = feb,
       volume = {135},
       number = {1044},
          eid = {028001},
        pages = {028001},
          doi = {10.1088/1538-3873/acac53},
archivePrefix = {arXiv},
       eprint = {2212.12069},
 primaryClass = {astro-ph.IM},
       adsurl = {https://ui.adsabs.harvard.edu/abs/2023PASP..135b8001R}
}

@ARTICLE{2024A&A...688A.180S,
       author = {{Scalco}, M. and {Bedin}, L. and {Vesperini}, E.},
        title = "{The HST Large Programme on {\ensuremath{\omega}} Centauri. VI. The radial gradient of the stellar populations}",
      journal = {\aap},
         year = 2024,
        month = aug,
       volume = {688},
          eid = {A180},
        pages = {A180},
          doi = {10.1051/0004-6361/202450344},
archivePrefix = {arXiv},
       eprint = {2405.19939},
 primaryClass = {astro-ph.GA},
       adsurl = {https://ui.adsabs.harvard.edu/abs/2024A&A...688A.180S}
}

@ARTICLE{2025A&A...694A..68S,
       author = {{Scalco}, M. and {Gerasimov}, R. and {Bedin}, L.~R. and {Vesperini}, E. and {Correnti}, M. and {Nardiello}, D. and {Burgasser}, A. and {Richer}, H. and {Caiazzo}, I. and {Heyl}, J. and {Libralato}, M. and {Anderson}, J. and {Griggio}, M.},
        title = "{JWST photometry and astrometry of 47 Tucan{\ae}: Discontinuity in the stellar sequence at the star{\textendash}brown dwarf transition}",
      journal = {\aap},
         year = 2025,
        month = feb,
       volume = {694},
          eid = {A68},
        pages = {A68},
          doi = {10.1051/0004-6361/202452907},
archivePrefix = {arXiv},
       eprint = {2501.04446},
 primaryClass = {astro-ph.SR},
       adsurl = {https://ui.adsabs.harvard.edu/abs/2025A&A...694A..68S}
}

@ARTICLE{MCMC,
       author = {{Goodman}, Jonathan and {Weare}, Jonathan},
        title = "{Ensemble samplers with affine invariance}",
      journal = {Communications in AMCS},
         year = 2010,
        month = jan,
       volume = {5},
       number = {1},
        pages = {65-80},
          doi = {10.2140/camcos.2010.5.65},
       adsurl = {https://ui.adsabs.harvard.edu/abs/2010CAMCS...5...65G}
}

@ARTICLE{2009ApJ...697L..58A,
       author = {{Anderson}, J. and {Piotto}, G. and {King}, I.~R. and {Bedin}, L.~R. and {Guhathakurta}, P. and {A}, A.},
        title = "{Mixed Populations in Globular Clusters: Et Tu, 47 Tuc?}",
      journal = {\apjl},
         year = 2009,
        month = may,
       volume = {697},
       number = {1},
        pages = {L58-L62},
          doi = {10.1088/0004-637X/697/1/L58},
archivePrefix = {arXiv},
       eprint = {0904.1626},
 primaryClass = {astro-ph.GA},
       adsurl = {https://ui.adsabs.harvard.edu/abs/2009ApJ...697L..58A}
}

@ARTICLE{2013ApJ...771L..15R,
       author = {{Richer}, Harvey B. and {Heyl}, Jeremy and {Anderson}, Jay and {Kalirai}, Jason S. and {Shara}, Michael M. and {Dotter}, Aaron and {Fahlman}, Gregory G. and {Rich}, R. Michael},
        title = "{A Dynamical Signature of Multiple Stellar Populations in 47 Tucanae}",
      journal = {\apjl},
         year = 2013,
        month = jul,
       volume = {771},
       number = {1},
          eid = {L15},
        pages = {L15},
          doi = {10.1088/2041-8205/771/1/L15},
archivePrefix = {arXiv},
       eprint = {1306.1226},
 primaryClass = {astro-ph.SR},
       adsurl = {https://ui.adsabs.harvard.edu/abs/2013ApJ...771L..15R}
}

@ARTICLE{emcee_2013,
       author = {{Foreman-Mackey}, Daniel and {Hogg}, David W. and {Lang}, Dustin and {Goodman}, Jonathan and {Goodman}, Jonathan and {Goodman}, Jonathan},
        title = "{emcee: The MCMC Hammer}",
      journal = {\pasp},
         year = 2013,
        month = mar,
       volume = {125},
       number = {925},
        pages = {306},
          doi = {10.1086/670067},
archivePrefix = {arXiv},
       eprint = {1202.3665},
 primaryClass = {astro-ph.IM},
       adsurl = {https://ui.adsabs.harvard.edu/abs/2013PASP..125..306F}
}

@ARTICLE{convective_kissing,
       author = {{van Saders}, Jennifer L. and {Pinsonneault}, Marc H.},
        title = "{An $^{3}$He-driven Instability near the Fully Convective Boundary}",
      journal = {\apj},
         year = 2012,
        month = jun,
       volume = {751},
       number = {2},
          eid = {98},
        pages = {98},
          doi = {10.1088/0004-637X/751/2/98},
archivePrefix = {arXiv},
       eprint = {1201.5381},
 primaryClass = {astro-ph.SR},
       adsurl = {https://ui.adsabs.harvard.edu/abs/2012ApJ...751...98V}
}

@ARTICLE{2021A&A...650A.184M,
       author = {{Mansfield}, Santana and {Kroupa}, Pavel},
        title = "{A discontinuity in the luminosity-mass relation and fluctuations in the evolutionary tracks of low-mass and low-metallicity stars at the Gaia M-dwarf gap}",
      journal = {\aap},
         year = 2021,
        month = jun,
       volume = {650},
          eid = {A184},
        pages = {A184},
          doi = {10.1051/0004-6361/202140536},
archivePrefix = {arXiv},
       eprint = {2104.13951},
 primaryClass = {astro-ph.SR},
       adsurl = {https://ui.adsabs.harvard.edu/abs/2021A&A...650A.184M}
}

@ARTICLE{2019A&A...629A..40S,
       author = {{Salaris}, M. and {Cassisi}, S. and {Mucciarelli}, A. and {Nardiello}, D.},
        title = "{Detection of multiple stellar populations in extragalactic massive clusters with JWST}",
      journal = {\aap},
         year = 2019,
        month = sep,
       volume = {629},
          eid = {A40},
        pages = {A40},
          doi = {10.1051/0004-6361/201936252},
archivePrefix = {arXiv},
       eprint = {1908.02229},
 primaryClass = {astro-ph.SR},
       adsurl = {https://ui.adsabs.harvard.edu/abs/2019A&A...629A..40S}
}

@ARTICLE{2018ApJ...861L..11J,
       author = {{Jao}, Wei-Chun and {Henry}, Todd J. and {Gies}, Douglas R. and {Hambly}, Nigel C.},
        title = "{A Gap in the Lower Main Sequence Revealed by Gaia Data Release 2}",
      journal = {\apjl},
         year = 2018,
        month = jul,
       volume = {861},
       number = {1},
          eid = {L11},
        pages = {L11},
          doi = {10.3847/2041-8213/aacdf6},
archivePrefix = {arXiv},
       eprint = {1806.07792},
 primaryClass = {astro-ph.SR},
       adsurl = {https://ui.adsabs.harvard.edu/abs/2018ApJ...861L..11J}
}

@ARTICLE{2026A&A...711A..88G,
       author = {{Griggio}, M. and {Libralato}, M. and {Gerasimov}, R. and {Bedin}, L.~R. and {Bellini}, A. and {Anderson}, J. and {Dalessandro}, E. and {Vesperini}, E. and {Baumgardt}, H. and {Massari}, D. and {Cadelano}, M. and {Ryan}, Jr., R.~E. and {McDonald}, I. and {Annibali}, F. and {Balbinot}, E. and {Cuillandre}, J.-C. and {Erkal}, D. and {Ferguson}, A.~M.~N. and {Kluge}, M. and {Kuzma}, P.~B. and {Saifollahi}, T. and {Schirmer}, M. and {Voggel}, K. and {Altieri}, B. and {Andreon}, S. and {Auricchio}, N. and {Baccigalupi}, C. and {Baldi}, M. and {Balestra}, A. and {Bardelli}, S. and {Battaglia}, P. and {Biviano}, A. and {Branchini}, E. and {Brescia}, M. and {Camera}, S. and {Capobianco}, V. and {Carbone}, C. and {Carretero}, J. and {Castellano}, M. and {Castignani}, G. and {Cavuoti}, S. and {Cimatti}, A. and {Colodro-Conde}, C. and {Congedo}, G. and {Conselice}, C.~J. and {Conversi}, L. and {Copin}, Y. and {Courbin}, F. and {Courtois}, H.~M. and {Cropper}, M. and {De Lucia}, G. and {Dubath}, F. and {Dupac}, X. and {Farina}, M. and {Farinelli}, R. and {Ferriol}, S. and {Franceschi}, E. and {Galeotta}, S. and {George}, K. and {Giocoli}, C. and {Gracia-Carpio}, J. and {Grazian}, A. and {Grupp}, F. and {Haugan}, S.~V.~H. and {Hoekstra}, H. and {Holmes}, W. and {Hormuth}, F. and {Hornstrup}, A. and {Jahnke}, K. and {Jhabvala}, M. and {Kermiche}, S. and {Kiessling}, A. and {Kubik}, B. and {K{\"u}mmel}, M. and {Kurki-Suonio}, H. and {Le Brun}, A.~M.~C. and {Ligori}, S. and {Lilje}, P.~B. and {Lindholm}, V. and {Lloro}, I. and {Mainetti}, G. and {Marggraf}, O. and {Martinelli}, M. and {Martinet}, N. and {Marulli}, F. and {Massey}, R.~J. and {Medinaceli}, E. and {Mei}, S. and {Meneghetti}, M. and {Merlin}, E. and {Meylan}, G. and {Mora}, A. and {Moresco}, M. and {Moscardini}, L. and {Nakajima}, R. and {Niemi}, S.-M. and {Padilla}, C. and {Paltani}, S. and {Pasian}, F. and {Pedersen}, K. and {Percival}, W.~J. and {Pettorino}, V. and {Polenta}, G. and {Poncet}, M. and {Popa}, L.~A. and {Raison}, F. and {Renzi}, A. and {Rhodes}, J. and {Riccio}, G. and {Romelli}, E. and {Roncarelli}, M. and {Saglia}, R. and {Sakr}, Z. and {Sapone}, D. and {Sartoris}, B. and {Schneider}, P. and {Secroun}, A. and {Simon}, P. and {Sirignano}, C. and {Sirri}, G. and {Stanco}, L. and {Tallada-Cresp{\'\i}}, P. and {Toft}, S. and {Toledo-Moreo}, R. and {Torradeflot}, F. and {Tutusaus}, I. and {Valiviita}, J. and {Vassallo}, T. and {Wang}, Y. and {Weller}, J. and {Zerbi}, F.~M. and {Zucca}, E. and {Scottez}, V.},
        title = "{Euclid: Early Release Observations ─ Internal kinematics and the convective-transition gap of NGC 6397: High-precision multiple-pass photometry and astrometry}",
      journal = {\aap},
         year = 2026,
        month = jul,
       volume = {711},
          eid = {A88},
        pages = {A88},
          doi = {10.1051/0004-6361/202660441},
archivePrefix = {arXiv},
       eprint = {2606.02703},
 primaryClass = {astro-ph.GA},
       adsurl = {https://ui.adsabs.harvard.edu/abs/2026A&A...711A..88G}
}

@ARTICLE{2018A&A...619A.177B,
       author = {{Baraffe}, Isabelle and {Chabrier}, Gilles},
        title = "{A closer look at the transition between fully convective and partly radiative low-mass stars}",
      journal = {\aap},
         year = 2018,
        month = nov,
       volume = {619},
          eid = {A177},
        pages = {A177},
          doi = {10.1051/0004-6361/201834062},
archivePrefix = {arXiv},
       eprint = {1809.07274},
 primaryClass = {astro-ph.SR},
       adsurl = {https://ui.adsabs.harvard.edu/abs/2018A&A...619A.177B}
}

@ARTICLE{2021ApJ...907...53F,
       author = {{Feiden}, Gregory A. and {Skidmore}, Khian and {Jao}, Wei-Chun},
        title = "{Gaia Gaps and the Physics of Low-mass Stars. I. The Fully Convective Boundary}",
      journal = {\apj},
         year = 2021,
        month = jan,
       volume = {907},
       number = {1},
          eid = {53},
        pages = {53},
          doi = {10.3847/1538-4357/abcc03},
archivePrefix = {arXiv},
       eprint = {2011.07991},
 primaryClass = {astro-ph.SR},
       adsurl = {https://ui.adsabs.harvard.edu/abs/2021ApJ...907...53F}
}

@ARTICLE{2020AJ....160..102J,
       author = {{Jao}, Wei-Chun and {Feiden}, Gregory A.},
        title = "{Fine Structures in the Main Sequence Revealed by Gaia Data Release 2}",
      journal = {\aj},
         year = 2020,
        month = sep,
       volume = {160},
       number = {3},
          eid = {102},
        pages = {102},
          doi = {10.3847/1538-3881/aba192},
       adsurl = {https://ui.adsabs.harvard.edu/abs/2020AJ....160..102J}
}

@ARTICLE{2018MNRAS.480.1711M,
       author = {{MacDonald}, James and {Gizis}, John},
        title = "{An explanation for the gap in the Gaia HRD for M dwarfs}",
      journal = {\mnras},
         year = 2018,
        month = oct,
       volume = {480},
       number = {2},
        pages = {1711-1714},
          doi = {10.1093/mnras/sty1888},
archivePrefix = {arXiv},
       eprint = {1806.11454},
 primaryClass = {astro-ph.SR},
       adsurl = {https://ui.adsabs.harvard.edu/abs/2018MNRAS.480.1711M}
}

@ARTICLE{2023ApJ...944..129B,
       author = {{Boudreaux}, Emily M. and {Chaboyer}, Brian C.},
        title = "{Updated High-temperature Opacities for the Dartmouth Stellar Evolution Program and Their Effect on the Jao Gap Location}",
      journal = {\apj},
         year = 2023,
        month = feb,
       volume = {944},
       number = {2},
          eid = {129},
        pages = {129},
          doi = {10.3847/1538-4357/acb685},
archivePrefix = {arXiv},
       eprint = {2301.10798},
 primaryClass = {astro-ph.SR},
       adsurl = {https://ui.adsabs.harvard.edu/abs/2023ApJ...944..129B}
}

@ARTICLE{2026arXiv260628504L,
       author = {{Libralato}, M. and {Griggio}, M. and {Gerasimov}, R. and {Bedin}, L.~R. and {McDonald}, I. and {Altieri}, B. and {Anderson}, J. and {Annibali}, F. and {Balbinot}, E. and {Battaglia}, G. and {Bellini}, A. and {Casenove}, P. and {Cuillandre}, J.-C. and {Dalessandro}, E. and {Ferguson}, A.~M.~N. and {Jones}, H.~R.~A. and {Kuijken}, K. and {Majidi}, F.~Z. and {Massari}, D. and {Mohandasan}, A. and {Niederhofer}, F. and {Polenta}, G. and {Sakowska}, J.~D. and {Soldano}, F. and {Zerbinati}, C. and {Andreon}, S. and {Auricchio}, N. and {Aussel}, H. and {Baccigalupi}, C. and {Baldi}, M. and {Balestra}, A. and {Battaglia}, P. and {Biviano}, A. and {Branchini}, E. and {Brescia}, M. and {Camera}, S. and {Capobianco}, V. and {Carbone}, C. and {Carretero}, J. and {Castellano}, M. and {Castignani}, G. and {Cavuoti}, S. and {Chambers}, K.~C. and {Cimatti}, A. and {Colodro-Conde}, C. and {Congedo}, G. and {Conselice}, C.~J. and {Conversi}, L. and {Copin}, Y. and {Courbin}, F. and {Courtois}, H.~M. and {Cropper}, M. and {Degaudenzi}, H. and {De Lucia}, G. and {Dole}, H. and {Dubath}, F. and {Dupac}, X. and {Farina}, M. and {Farinelli}, R. and {Faustini}, F. and {Ferriol}, S. and {Frailis}, M. and {Franceschi}, E. and {Galeotta}, S. and {George}, K. and {Gillis}, B. and {Giocoli}, C. and {Gracia-Carpio}, J. and {Grazian}, A. and {Grupp}, F. and {Haugan}, S.~V.~H. and {Hoekstra}, H. and {Holmes}, W. and {Hook}, I.~M. and {Hormuth}, F. and {Hornstrup}, A. and {Jahnke}, K. and {Jhabvala}, M. and {Kermiche}, S. and {Kiessling}, A. and {Kubik}, B. and {K{\"u}mmel}, M. and {Kunz}, M. and {Kurki-Suonio}, H. and {Le Brun}, A.~M.~C. and {Ligori}, S. and {Lilje}, P.~B. and {Lindholm}, V. and {Lloro}, I. and {Mansutti}, O. and {Marggraf}, O. and {Martinelli}, M. and {Martinet}, N. and {Marulli}, F. and {Massey}, R.~J. and {Medinaceli}, E. and {Mei}, S. and {Meneghetti}, M. and {Merlin}, E. and {Meylan}, G. and {Mora}, A. and {Moscardini}, L. and {Nakajima}, R. and {Nichol}, R.~C. and {Niemi}, S.-M. and {Padilla}, C. and {Paltani}, S. and {Pasian}, F. and {Percival}, W.~J. and {Pettorino}, V. and {Poncet}, M. and {Popa}, L.~A. and {Raison}, F. and {Renzi}, A. and {Rhodes}, J. and {Riccio}, G. and {Rizzo}, F. and {Romelli}, E. and {Roncarelli}, M. and {Saglia}, R. and {Sakr}, Z. and {Sapone}, D. and {Schirmer}, M. and {Schneider}, P. and {Schrabback}, T. and {Secroun}, A. and {Sihvola}, E. and {Simon}, P. and {Sirignano}, C. and {Sirri}, G. and {Stanco}, L. and {Tallada-Cresp{\'\i}}, P. and {Taylor}, A.~N. and {Tereno}, I. and {Toft}, S. and {Toledo-Moreo}, R. and {Torradeflot}, F. and {Tutusaus}, I. and {Valiviita}, J. and {Vassallo}, T. and {Wang}, Y. and {Weller}, J. and {Scott}, D.},
        title = "{Euclid: The convective-transition gap of 47 Tuc}",
      journal = {arXiv e-prints},
         year = 2026,
        month = jun,
          eid = {arXiv:2606.28504},
        pages = {arXiv:2606.28504},
          doi = {10.48550/arXiv.2606.28504},
archivePrefix = {arXiv},
       eprint = {2606.28504},
 primaryClass = {astro-ph.SR},
       adsurl = {https://ui.adsabs.harvard.edu/abs/2026arXiv260628504L}
}

@ARTICLE{2021ApJ...908..102P,
       author = {{Pietrinferni}, Adriano and {Hidalgo}, Sebastian and {Cassisi}, Santi and {Salaris}, Maurizio and {Savino}, Alessandro and {Mucciarelli}, Alessio and {Verma}, Kuldeep and {Silva Aguirre}, Victor and {Aparicio}, Antonio and {Ferguson}, Jason W.},
        title = "{Updated BaSTI Stellar Evolution Models and Isochrones. II. {\ensuremath{\alpha}}-enhanced Calculations}",
      journal = {\apj},
         year = 2021,
        month = feb,
       volume = {908},
       number = {1},
          eid = {102},
        pages = {102},
          doi = {10.3847/1538-4357/abd4d5},
archivePrefix = {arXiv},
       eprint = {2012.10085},
 primaryClass = {astro-ph.SR},
       adsurl = {https://ui.adsabs.harvard.edu/abs/2021ApJ...908..102P}
}

\begin{appendix}

\section{Dataset}\label{Appendix:A}

\begin{table*}[h!]
\centering
\begin{threeparttable}
\caption{Summary of the dataset used in this work, sorted by increasing filter effective wavelength.}
\label{table:dataset}

\onecolumn
\begin{tabular*}{\textwidth}{@{\extracolsep{\fill}}l c c c c c}
\hline
\hline
Filter & Instrument & $N \times$ Exp. time & Date & Programme ID & PI\\
\hline

\multicolumn{6}{c}{JWST}\\
\hline

F070W  & NIRCam & $2\times1,611$\,s & Aug. 16--27, 2024 & 6631 & Koekemoer\\
F070W  & NIRCam & $9\times236$\,s & Oct. 22, 2025 & 8816 & Rest\\
F090W  & NIRCam & $2\times1,611$\,s & Aug. 16--27, 2024 & 6631 & Koekemoer\\
F090W  & NIRCam & $6\times623$\,s & Sep.  8, 2025 & 5896 & Scalco\\
F115W  & NIRCam & $9\times236$\,s & Oct. 22, 2025 & 8816 & Rest\\
F150W  & NIRCam & $2\times1,074$\,s & Aug. 16--27, 2024 & 6631 & Koekemoer\\
F150W2 & NIRCam & $48\times837$\,s & Sep. 14--15, 2022 & 2559 & Caiazzo\\
F150W2 & NIRCam & $6\times945$\,s & May 24--25, 2023 & 2559 & Caiazzo\\
F200W  & NIRCam & $9\times236$\,s & Oct. 22, 2025 & 8816 & Rest\\
F277W  & NIRCam & $2\times1,611$\,s & Aug. 16--27, 2024 & 6631 & Koekemoer\\
F277W  & NIRCam & $9\times236$\,s & Oct. 22, 2025 & 8816 & Rest\\
F322W2 & NIRCam & $48\times837$\,s & Sep. 14--15, 2022 & 2559 & Caiazzo\\
F322W2 & NIRCam & $6\times945$\,s & May 24--25, 2023 & 2559 & Caiazzo\\
F356W  & NIRCam & $2\times1,611$\,s & Aug. 16--27, 2024 & 6631 & Koekemoer\\
F356W  & NIRCam & $9\times236$\,s & Oct. 22, 2025 & 8816 & Rest\\
F444W  & NIRCam & $2\times1,074$\,s & Aug. 16--27, 2024 & 6631 & Koekemoer\\
F444W  & NIRCam & $6\times623$\,s & Sep.  8, 2025 & 5896 & Scalco\\
F444W  & NIRCam & $9\times236$\,s & Oct. 22, 2025 & 8816 & Rest\\

\hline
\end{tabular*}

\end{threeparttable}
\end{table*}

\begin{centering} 
\begin{figure}[h!]
 \centering
 \includegraphics[width=.675\columnwidth]{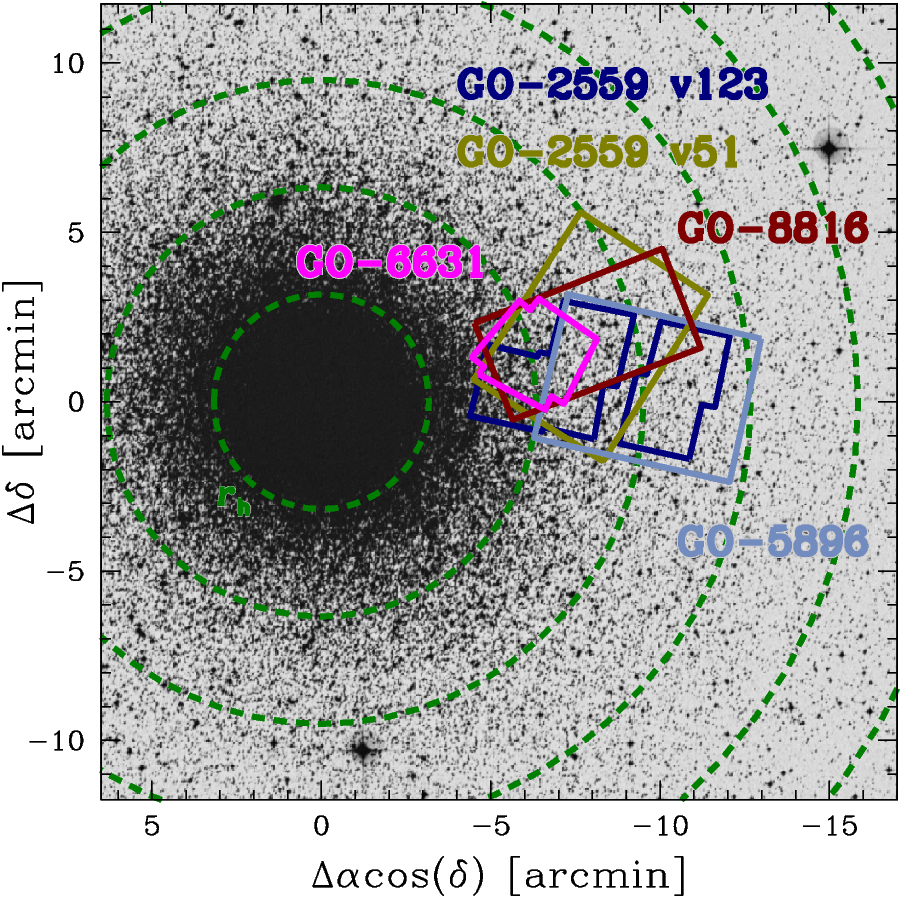}
 \caption{Footprints of the JWST observations used in this work overlaid on a Digital Sky Survey (DSS) image of 47\,Tuc. Different colours identify the individual observing programmes (GO-2559 visits v123 and v51, GO-6631, GO-5896, and GO-8816), as labelled in the figure. Axes are given in arcminutes relative to the cluster centre. The green dashed circles mark the cluster half-light radius, $r_{\rm h}=3\farcm17$ \citep{2010arXiv1012.3224H}, and its integer multiples. The combined dataset samples the cluster over a projected radial range of approximately $1.5 \lesssim R/r_{\rm h} \lesssim 4$.}
 \label{FOV} 
\end{figure} 
\end{centering}

\onecolumn
\section{Artificial-star tests}\label{Appendix:E}

To assess the effect of photometric uncertainties on the observed LF, we performed a series of AS tests, focusing on the F200W dataset. Artificial-star magnitudes were generated from a set of input LFs, while their positions were randomly distributed over the F200W FOV. Three different input LFs were considered: (i) a flat LF, (ii) a gap LF, corresponding to the best-fitting LF presented in panel\,(b) of Fig.\,\ref{fig:lf2}, and (iii) a gap-free LF corresponding to the same best-fitting LF, but with the gap removed ($D=1$). For each LF, we injected 5000 ASs uniformly distributed over the magnitude range $19.55<m_{\rm F200W}<20.55$, matching the number of real stars observed in the same magnitude interval.

The ASs were injected, detected, and measured following exactly the same procedures adopted for the real stars. An AS was considered successfully recovered if the input and output positions differed by less than one pixel, and the input and output magnitudes differed by less than 0.75\,mag (corresponding to $\sim2.5\log2$).

\begin{figure}[!h]
 \centering
 \includegraphics[width=0.88\columnwidth]{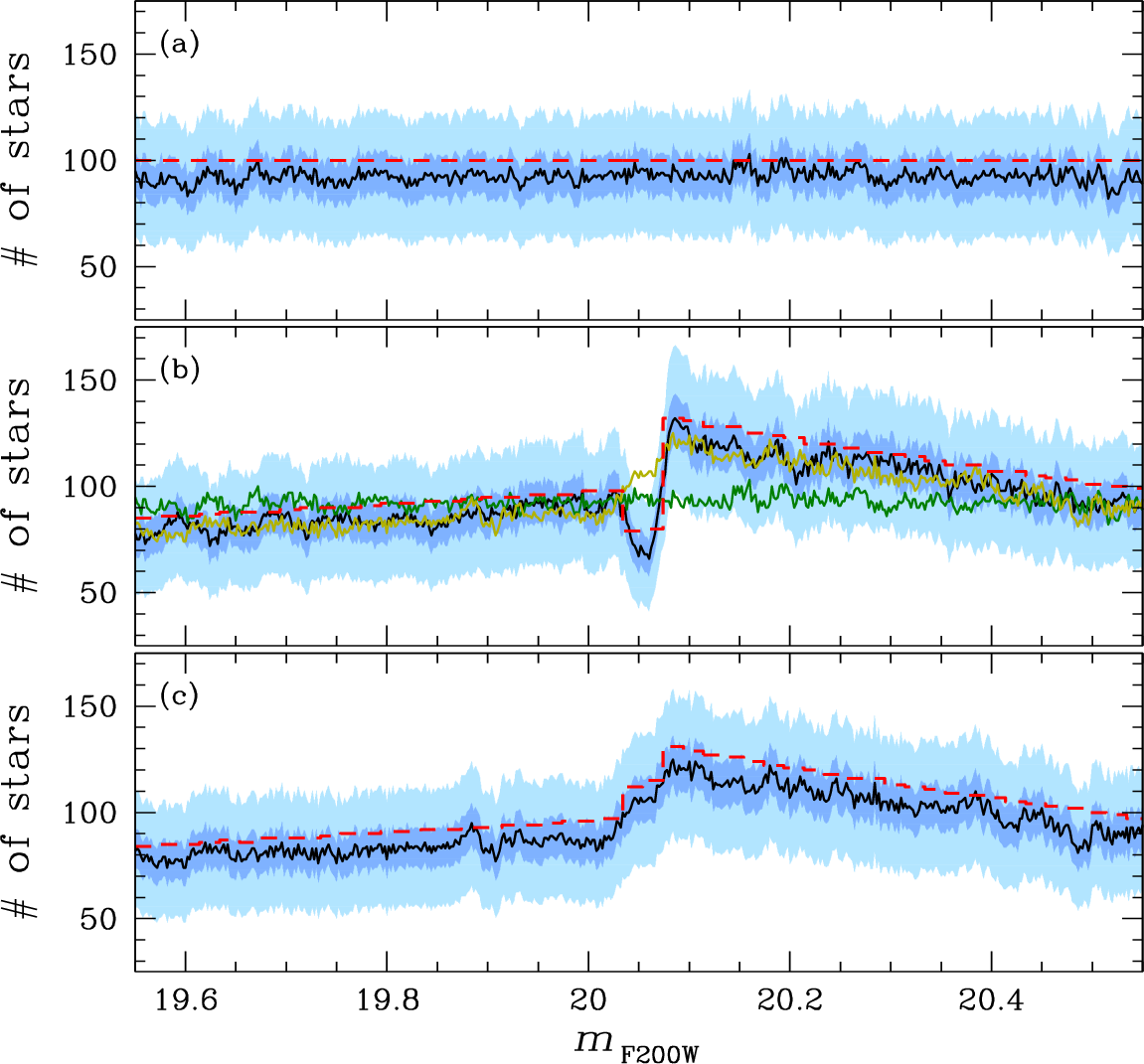}
 \caption{Artificial-star tests based on three different input LFs: a flat LF (a), a gap LF, corresponding to the best-fitting model from panel\,(b) of Fig.\,\ref{fig:lf2}, and the same LF as in (b), but with the gap removed ($D=1$; c). The three input LFs are shown in red. The recovered LFs are shown as black curves, with the dark- and light-blue shaded regions indicating the corresponding 1$\sigma$ and 3$\sigma$ confidence intervals, respectively. For comparison, the recovered flat and gap-free LFs shown in black in panels\,a and c are also overplotted in panel\,(b) in green and yellow, respectively.}
 \label{fig:ast} 
\end{figure}

The injected and recovered LFs are shown in Fig.\,\ref{fig:ast}. The recovered flat LF remains flat over the entire magnitude range (panel\,a), showing that the photometric uncertainties do not introduce artificial local features, while the gap LF preserves its characteristic morphology after recovery (panel\,b). Similarly, the gap-free LF does not develop any spurious local features after the recovery (panel\,c). A comparison between the recovered flat and gap-free LFs and the recovered gap LF is presented in panel\,(b). In the vicinity of the gap, the recovered gap-free LF lies well outside the $3\sigma$ confidence region of the recovered gap LF, while the recovered flat LF reaches the boundary of the $3\sigma$ region at the gap and at the subsequent step. Fitting the recovered gap LF with the same maximum-likelihood procedure adopted for the observed data yields $\Delta \ln \mathcal{L} = \ln \mathcal{L}_{\rm gap} - \ln \mathcal{L}_{\rm no\,gap} = 15.55$, confirming the preference for the gap model.
 
\onecolumn
\section{Monte Carlo posterior distributions}\label{Appendix:B}
 
\begin{figure*}[h!]
 \centering
 \includegraphics[width=0.75\textwidth]{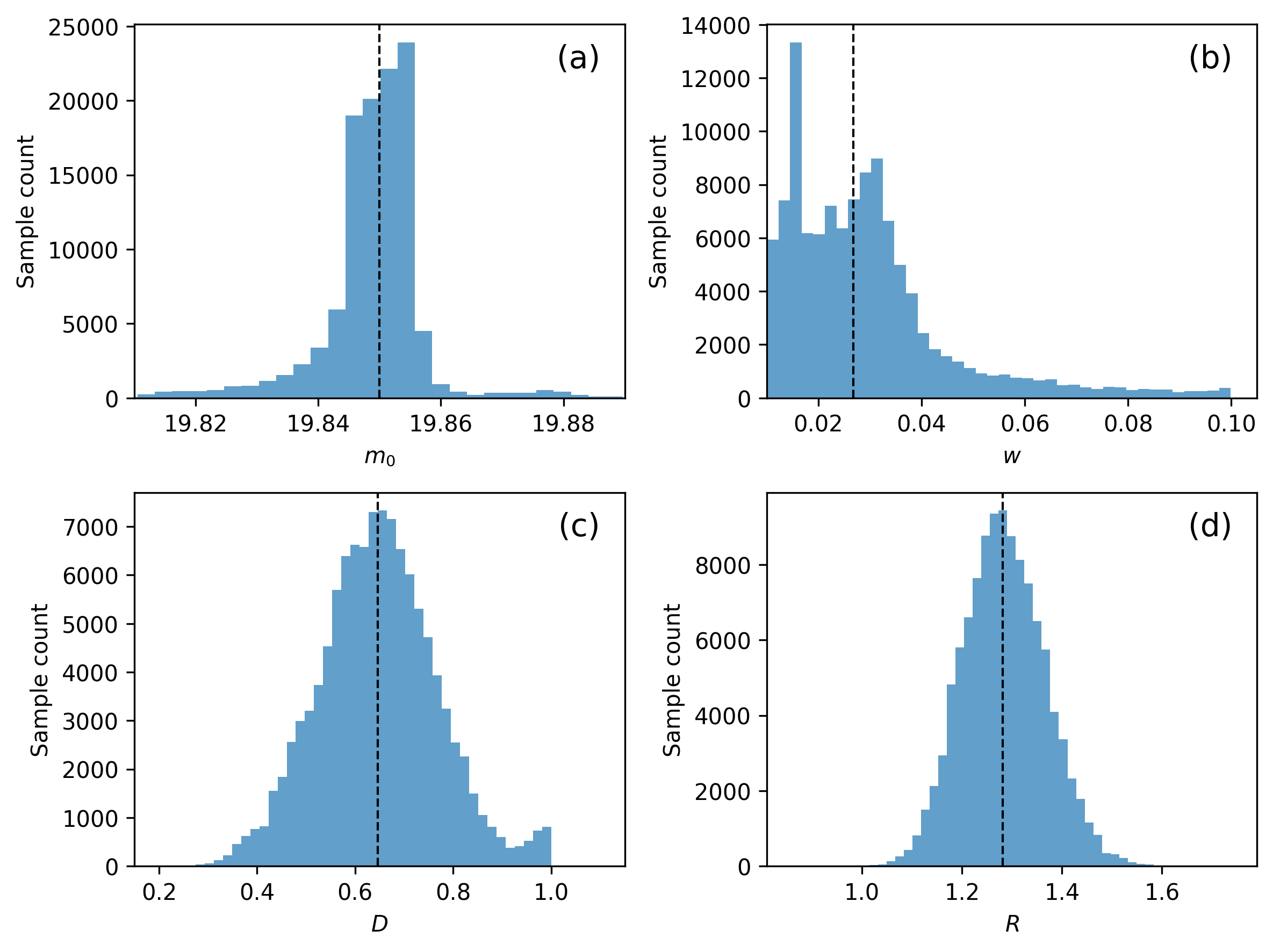}
 \caption{Posterior distributions of the gap parameters for the $m_{\rm F322W2}$ magnitude.} 
 \label{posterior322} 
\end{figure*} 

\begin{figure*}[h!]
\centering
 \includegraphics[width=0.75\textwidth]{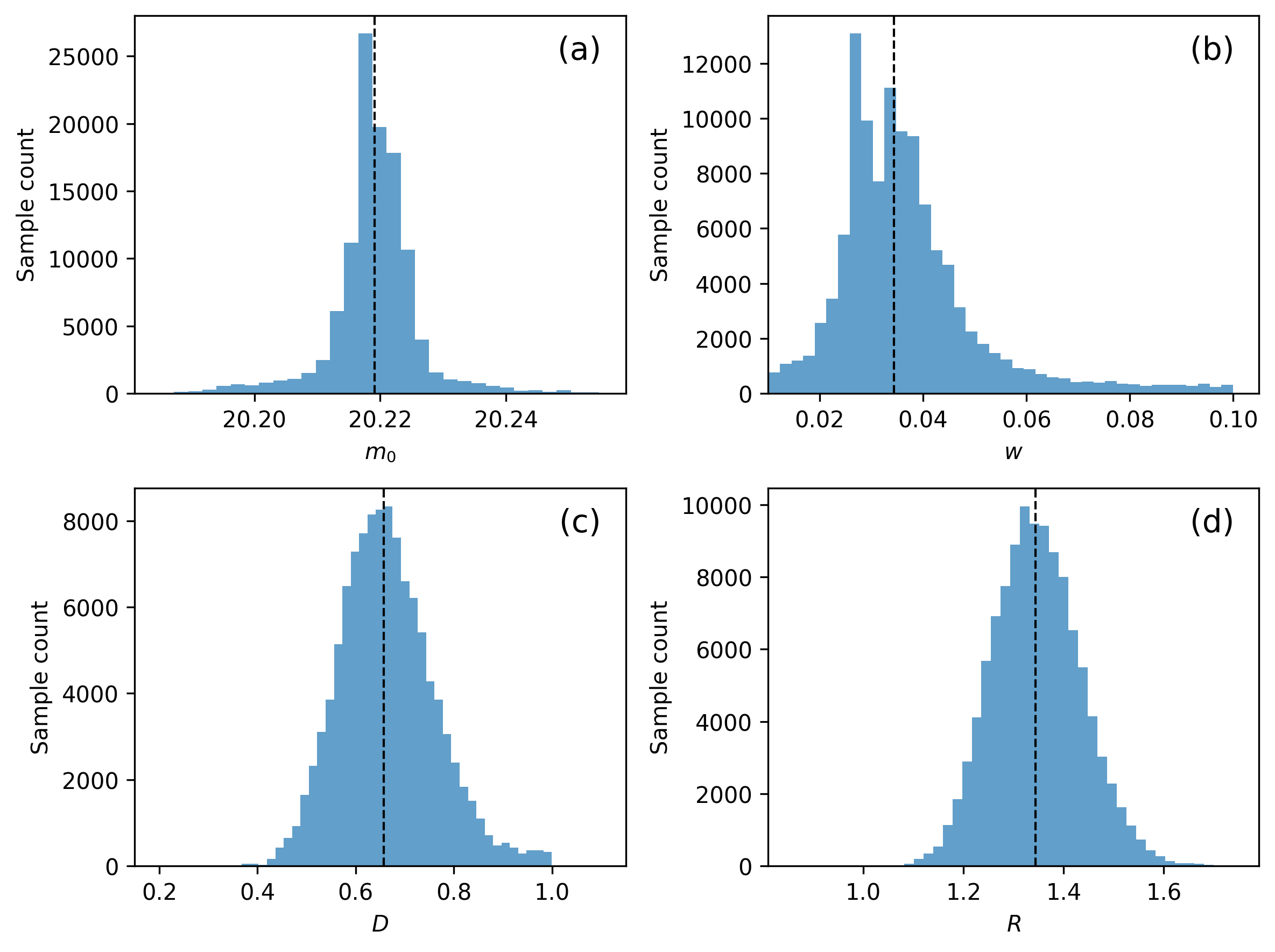}
 \caption{Same as Fig.\,\ref{posterior322} but for the $m_{\rm F200W}$ magnitude.} 
 \label{posterior200} 
\end{figure*} 

\onecolumn
\section{Additional luminosity functions}\label{Appendix:C}

\begin{figure}[!h]
    \centering
    \includegraphics[width=\columnwidth]{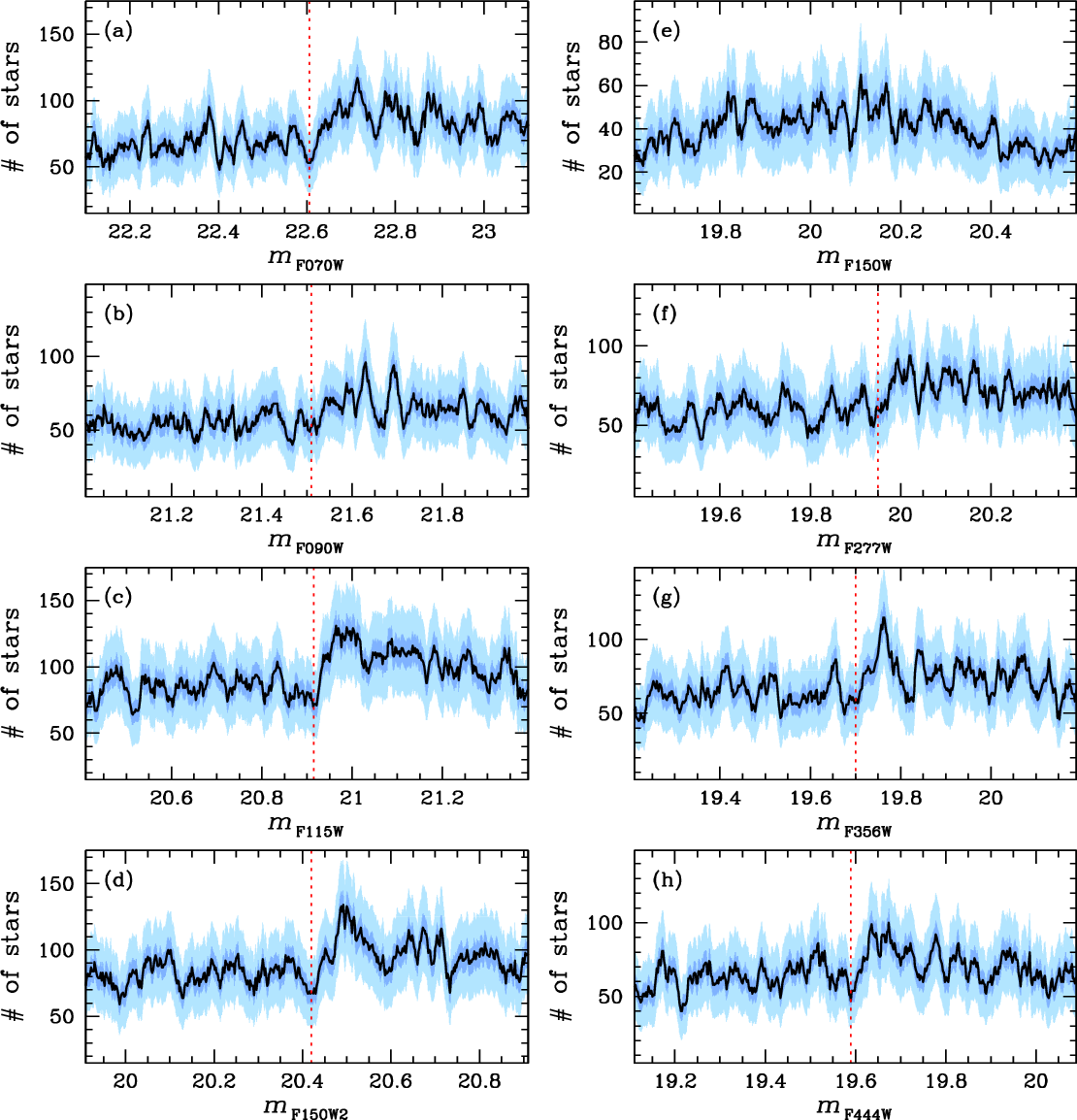} 
    \caption{Luminosity functions of 47\,Tuc in eight JWST/NIRCam filters around the convective-transition region. The vertical red dashed lines indicate the tentative location of the discontinuity associated with the convective-transition gap. The visibility of the feature varies among filters: a local deficit of stars followed by a discontinuous increase in the stellar counts is visible in the F070W and F150W2 LFs, while a pronounced discontinuity is apparent in F115W. Similar, although less significant, features are visible in the F090W, F277W, F356W, and F444W LFs. No clear signature of the gap is detected in F150W, likely owing to the limited number of stars available in this filter, resulting from the smaller area covered by the GO-6631 observations, whereas the other datasets sample broadly comparable regions of the cluster.}
    \label{ist_tot}
\end{figure}

\onecolumn
\section{Two-pseudo-colour diagrams and population separation}\label{Appendix:D}

\begin{figure}[!h]
    \centering
    \includegraphics[width=\columnwidth]{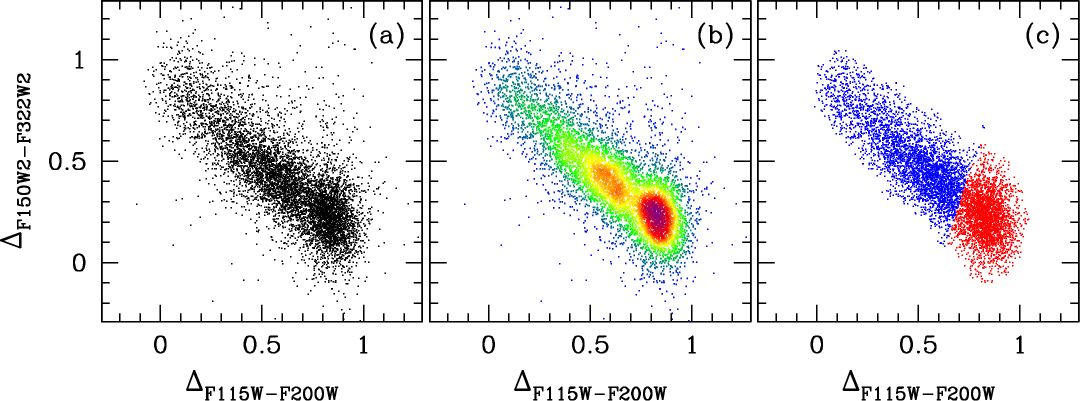}    
    \caption{Two-pseudo-colour diagram (TpCD) used to separate the two main stellar populations of 47\,Tuc. (a) Distribution of stars in the $m_{\rm F150W2}-m_{\rm F322W2}$ versus $m_{\rm F115W}-m_{\rm F200W}$ TpCD (b) Density map of the TpCD. The diagram reveals two main overdensities: a prominent clump in the lower-right region, associated with the oxygen-rich 1P, and a broader structure extending towards the upper-left region, associated with the oxygen-poor 2P. The separation is driven primarily by differences in the $m_{\rm F115W}-m_{\rm F200W}$ colour, which is highly sensitive to oxygen abundance variations in very low-mass stars owing to the strong H$_2$O molecular absorption bands sampled by the F200W filter \citep[see e.g.][]{2019A&A...629A..40S}. (c) Selected samples of stars, with the two populations, 1P and 2P, highlighted in red and blue, respectively.}
    \label{tpcd}
\end{figure}

\end{appendix}

\end{document}